\documentclass{article}

\usepackage{arxiv}

\usepackage[utf8]{inputenc} 
\usepackage[T1]{fontenc}    
\usepackage{hyperref}       
\usepackage{url}            
\usepackage{booktabs}       
\usepackage{amsfonts}       
\usepackage{nicefrac}       
\usepackage{microtype}      
\usepackage{amsmath}    
\usepackage{lipsum}
\usepackage{graphicx}
\graphicspath{ {./images/} }

\title{Braided interferometer mesh for robust photonic matrix-vector multiplications with non-ideal components}

\author{
 Federico Marchesin \\
 Photonics Research Group\\
 Department of Information Technology\\
 Ghent University-imec, Ghent, Belgium\\
 \texttt{Federico.Marchesin@UGent.be} \\
\AND
 Matěj Hejda \\
 Hewlett Packard Labs, Diegem, Belgium\\
\And
 Tzamn Melendez Carmona \\
 Department of Control and Computer Engineering,\\
 Politecnico di Torino, Turin, Italy\\
\And
 Stefano Di Carlo \\
 Department of Control and Computer Engineering,\\
 Politecnico di Torino, Turin, Italy\\
\And
 Alessandro Savino \\
 Department of Control and Computer Engineering,\\
 Politecnico di Torino, Turin, Italy\\
\And
 Fabio Pavanello \\
 Univ. Grenoble Alpes, Univ. Savoie Mont Blanc, CNRS,\\
 Grenoble INP, CROMA, Grenoble, France\\
\And
 Thomas Van Vaerenbergh \\
 Hewlett Packard Labs, Diegem, Belgium\\
\And
 Peter Bienstman \\
 Photonics Research Group,\\
 Department of Information Technology,\\
 Ghent University-imec, Ghent, Belgium\\
}

\begin{document}
\maketitle
\begin{abstract}
Matrix-vector multiplications (MVMs) are essential for a wide range of applications, particularly in modern machine learning and quantum computing. In photonics, there is growing interest in developing architectures capable of performing linear operations with high speed, low latency, and minimal loss. Traditional interferometric photonic architectures, such as the Clements design, have been extensively used for MVM operations. However, as these architectures scale, improving stability and robustness becomes critical. In this paper, we introduce a novel photonic braid interferometer architecture that outperforms both the Clements and Fldzhyan designs in these aspects. Using numerical simulations, we evaluate the performance of these architectures under ideal conditions and systematically introduce non-idealities such as insertion losses, beam splitter imbalances, and crosstalk. The results demonstrate that the braid architecture offers superior robustness due to its symmetrical design and reduced layer count.
Further analysis shows that the braid architecture is particularly advantageous in large-scale implementations, delivering better performance as the size of the interferometer increases. We also assess the footprint and total insertion losses of each architecture. Although waveguide crossings in the braid architecture slightly increase the footprint and insertion loss, recent advances in crossing technology significantly minimize these effects. Our study suggests that the braid architecture is a robust solution for photonic neuromorphic computing, maintaining high fidelity in realistic conditions where imperfections are inevitable.
\end{abstract}


\section{Introduction}

Linear algebraic operations such as general matrix-matrix multiplication (GeMM) and matrix-vector multiplications (MVMs) represent one of the fundamental low-level operations in many ubiquitous algorithms, including solvers of linear systems, optimization problems, or machine learning algorithms. Given the explosive growth of modern artificial intelligence (AI) in recent years and the corresponding increase in the scale and complexity of the AI model \cite{Sevilla2022_2IJCNNI}, the ability to effectively compute and accelerate such operations is critical. Typically, these operations are accelerated in dedicated hardware coprocessors with a high degree of parallelism, such as graphical processing units (GPUs) or tensor processing units (TPUs) \cite{Jouppi2023_2ISCA}. Furthermore, the development of novel computing substrates for AI acceleration is a thriving research field, with some of the leading technologies including memristors \cite{Dalgaty2024_NatComm}, spintronics \cite{Ross2023_NatNano} and photonics \cite{Shastri2021_NatPhot}.

Thanks to its use of optical signals, photonics offers a multitude of highly desirable properties for computing hardware. These include the mutually noninteracting nature of lightwaves, the feasibility of ultra-low-loss signal waveguiding, signal propagation without Joule heating, and the feasibility of using additional degrees of freedom for representing or multiplexing information, such as wavelength, phase, or polarization. In photonic integrated circuits (PICs), MVMs are typically realized by implementing the matrix in the programmable photonic circuit state \cite{Bogaerts2020_Nature}, and the vectors are encoded in the input lightwaves. The two most commonly studied types of MVM PICs are multiport interferometer meshes and photonic crossbars \cite{Feldmann2021_Nature, giamougiannis_coherent_2023}. In particular, integrated interferometric meshes are well suited for coherent MVM engines and implement unitary transform matrices (under the assumption of lossless circuits). Some reference designs include the triangular Mach-Zehnder interferometer mesh (MZI), for which Reck proposed a phase decomposition algorithm in \cite{Reck1994_PRL}. This design was used in the seminal work \cite{Shen2017_NatPhot} demonstrating the suitability of chained MZI meshes for the realization of arbitrary linear transforms by singular value decomposition (SVD), allowing the linear layers of the optical neural network (ONN). Another reference design is the rectangular MZI mesh, which reduces the mesh depth and provides more circuit symmetry. Clements proposed a phase decomposition algorithm for this case in \cite{clements_optimal_2016}. Some of the alternative MZI mesh designs include the Bokun mesh \cite{Mojaver2023_OE}, the diamond mesh \cite{Hamerly2024_X} and the path-independent loss (PILOSS) circuit \cite{Suzuki2019_JLT}. If we generalize beyond MZI building blocks, generic multiport unitary interferometer circuits with lumped elements also include the FFTnet \cite{Feng2022_ACSPhotonics} (which is related to the Beneš routing network \cite{Lenfant1978_IToC}) or the Fldzhyan design \cite{fldzhyan_optimal_2020}.

In this paper, we propose a novel unitary interferometer architecture, termed braid architecture, which addresses many shortcomings in existing architectures, such as the Clements and Fldzhyan architectures. The braid architecture offers enhanced robustness against non-idealities, delivering higher fidelity of the expressed matrix even when operated with imperfect components. Through comprehensive simulations and performance evaluations, we demonstrate that this new design outperforms current state-of-the-art architectures, positioning it as a promising candidate for future photonic MVM acceleration technologies.

This work presents an in-depth analysis of the most critical non-idealities that influence the performance of photonic SVD-based architectures. By examining their effects on fidelity, we aim to understand better limitations and opportunities for operation under real-world, non-ideal conditions. Our findings will contribute valuable information on the practical deployment of photonic neural networks and photonic quantum applications \cite{de_marinis_photonic_2019, peserico_integrated_2023, cheng_photonic_2021}, where minimizing the impact of component non-idealities is crucial for achieving high-performance results.

The remainder of this paper is structured as follows. In Section \ref{sec:theory}, we review the theory behind planar interferometers for linear components, describe the basic components, and describe how to arrange them in the braid architecture. In Section \ref{sec:FidelitySimulation}, we explain the methodology we used to evaluate the fidelity of the different architectures. The main part of this paper is Section \ref{sec:fidelity_analisys}, which presents the simulation results for different component non-idealities. In addition, we consider other key factors, such as the physical footprint (Section \ref{footprint}) and the loss of insertion (Section \ref{insertion_loss}) of each architecture. These factors play a significant role in determining photonic circuits' scalability and practical deployment, especially since they are integrated into larger systems.


\section{Theory of planar interferometers for linear operations}
\label{sec:theory}

Any $N$ port interferometer can be represented by a transfer matrix $N \times N$ with complex values $M$ in the frequency domain, which transforms an input electric field vector into an output vector, following the relation $E_{out} = M E_{in}$.

The most commonly applied technique to realize universal matrices in photonic integrated architectures is based on Singular Value Decomposition (SVD). SVD is a fundamental mathematical method that factorizes any complex matrix $m \times n$ $M$ into three distinct matrices as $M = U \Sigma V^\dagger$. In this factorization, $U$ is a $m \times m$ unitary matrix, $\Sigma$ is a $m \times n$ diagonal matrix, and $V^\dagger$ represents the conjugate transpose of $V$, where $V$ is an $n \times n$ unitary matrix. In integrated photonics, the diagonal matrix $\Sigma$ can be implemented using an array of Mach-Zehnder modulators (MZM), where we assume that the gain can also be present in the MZM to avoid reducing the matrix expressivity. In contrast, the $U$ and $V^\dagger$ matrices require a more complex photonic interferometric architecture.

In this paper, we will consider different decompositions of unitary matrices as a cascade of several layers. Each layer consists of several basic photonic components operating in parallel, i.e., independently within that layer. The general matrix $U$ can be represented as:

\begin{equation}
U = D \left( \prod_{i \in S} T_i \right)
\label{eq:unitaryMatrixDecomposition}
\end{equation}

where $D$ is a diagonal matrix that modulates the phase of the complex signal, while $T_i$ represents the transfer function of the individual layers in the cascade. The sequence $S$ represents the arrangement of the layers, with the leftmost layer appearing at the rightmost end of the product. Each layer within the architecture can, to a first approximation, be considered to have a transfer function independent of the others, under the approximation that reflections are neglected.

As mentioned, the diagonal matrix $D$ in Equation \ref{eq:unitaryMatrixDecomposition} allows control over the phase of each output and the total global phase. Without this matrix, it would not be possible to fully represent the $U(n)$ (unitary group) space, as the individual output phases would be incorrect \cite{clements_optimal_2016, machoortiz_optical_2021}. When using SVD to reproduce any universal matrix, it is important to note that the diagonal matrix $D$ in the unitary architectures can be omitted. In fact, it can be shown that the diagonal matrices $D$ associated with the unitary matrices $U$ and $V$ can be effectively absorbed into the diagonal complex matrix $\Sigma$.

The layers that make up the three unitary architectures that we will discuss in this paper (Clements \cite{clements_optimal_2016}, Fldzhyan \cite{fldzhyan_optimal_2020}, and our proposed braid architecture) share the same basic optical components, specifically optical beam splitters (BS) and phase shifters (PhS). The braid architecture also includes waveguide crossings, whether they should be considered distinct components is debatable, as they merely represent intersections between waveguides. In our analysis, waveguides are assumed to be ideal optical connections, but we will address the impact of non-ideal crossings on the performance of the braid architecture.

In our formalism, we consider each layer to consist of several optical components that operate in parallel and independently from each other, i.e., on distinct sets of input channels. Thus, the layer transfer matrix $T_i$ can be represented as a block diagonal matrix:

\begin{equation}
T_i =
\begin{bmatrix}
TM_1 & 0 & \cdots & 0 \\
0 & TM_2 & \cdots & 0 \\
\vdots & \vdots & \ddots & \vdots \\
0 & 0 & \cdots & TM_n
\end{bmatrix}
\label{eq:blockMatrix}
\end{equation}

Here, $TM_j$ represents the different transfer matrices $2 \times 2$ or $1 \times 1$ of the individual photonic components.

\subsection{Detailed overview of photonic components}

One required component is 2x2 beam splitters (BS) with a 50/50 power splitting ratio (3dB splitters). These can be implemented using different designs, such as multimode ferometers (MMIs) or directional couplers (DC). MMIs offer advantages, including wavelength insensitivity, ease of fabrication, and versatility in power-splitting ratio. In contrast, DCs provide compact size and lower loss, but they are more sensitive to wavelength variations and involve greater fabrication complexity.

In general, the transfer matrix of a BS, accounting for insertion losses and imbalances, can be approximated as:

\begin{equation}
T_\mathrm{BS} = \sqrt{\mathrm{IL}_\mathrm{lin}}
\begin{bmatrix}
\sqrt{1/2+\alpha_{BS}} & j \sqrt{1/2-\alpha_{BS}} \\
j \sqrt{1/2-\alpha_{BS}} & \sqrt{1/2+\alpha_{BS}} \\
\end{bmatrix}
\end{equation}

Here, $\mathrm{IL}_\mathrm{lin}$ represents the linear insertion losses, we can define it as: $ \mathrm{IL}_\mathrm{{dB}} = -10\mathrm{log}_{10} \mathrm{IL}_\mathrm{lin} = -10\mathrm{log}_{10} \mathrm{P}_\mathrm{out}/\mathrm{P}_\mathrm{in} $. These are expressed in terms of power, so we apply a square root to express them in terms of the electric field. The parameter $\alpha_\mathrm{BS}$ represents the imbalance factor in the beam splitter, which is derived from the power imbalance. We can derive that $\alpha_\mathrm{BS} = \frac{1}{2} \frac{\mathrm{IMB}-1}{\mathrm{IMB}+1}$ where $\mathrm{IMB}$ is the linear power imbalance, calculated as the ratio of the maximum power $P_\mathrm{max}$ to the minimum power $P_\mathrm{min}$ observed in the two outputs of the beam splitter.

The second component, the phase shifter \cite{sun_silicon_2022}, is arguably the most important one, as it is needed to program the unitary matrix. There are several options for phase shifters, each with its advantages and disadvantages. Thermo-optic phase shifters \cite{liu_thermo-optic_2022} and Micro-Electro-Mechanical Systems (MEMS) \cite{quack_integrated_2023} offer solutions with very low power losses, typically below 0.5 dB, although this depends on the specific design and materials used. However, these advantages often come at the expense of slower response times. These are less suitable for applications where fast response is crucial. Moreover, maintaining the required phase shift in thermo-optic devices demands electrical power to control and sustain the necessary temperature, leading to high power consumption.

On the other hand, Electro-Optic phase shifters (EOPS) \cite{witzens_high-speed_2018} provide much higher response speeds, but they have higher insertion losses and a larger footprint. They exhibit insertion losses in the range of 1 to 3 dB, this can be higher in some cases depending on the material platform and fabrication process. Another emerging technology is phase-change materials (PCMs) \cite{matos_review_2023, fang_non-volatile_2023, cheung_ultrapowerefficient_2024}. When stimulated by heat or an electric field, PCMs are based on materials such as ternary chalcogenide compounds that can undergo reversible phase transitions (such as from amorphous to crystalline). These changes result in a significant alteration in the complex refractive index of the material, which can be used to control the intensity or phase of light in a waveguide. A key advantage of PCMs is their non-volatile character: once a material phase change has occurred, the new state remains stable without requiring continuous power supply. The drawback of PCMs is that their insertion losses are dependent on the material state. For GST, the insertion losses range from approximately 1-3 dB in the crystalline state, increasing to 2–5 dB in the amorphous state due to higher optical absorption. This variation in loss between states negatively impacts the overall performance of the architectures. Ideally, we would prefer PCMs that exhibit minimal losses \cite{delaney_nonvolatile_2021, dwivedi_ultra-low-loss_2024}, where only the phase is modulated without introducing significant attenuation of the amplitude.

The transfer function for the phase shifter can be written as follows:

\begin{equation}
T_\mathrm{PhS} = \sqrt{\mathrm{IL}_\mathrm{lin}} e^{j \theta}
\end{equation}

where $\mathrm{IL}_\mathrm{lin}$ accounts for the insertion losses and $\theta$ represents the phase shift introduced by the phase shifter.

The last component we analyze is the waveguide crossing \cite{johnson_low-loss_2020}. Among the architectures analyzed, it is present only in the braid architecture. Although many similar architectures have been proposed that use this component \cite{Feng2022_ACSPhotonics, gu_adept_2022}, none have been demonstrated to represent a fully unitary matrix architecture. This component is considered non-ideal because of its insertion loss and crosstalk, which can further degrade the circuit performance.

The transfer matrix for a non-ideal crossing can be expressed as:
\begin{equation}
T_{x} = \sqrt{\mathrm{IL}_\mathrm{lin}}
\begin{bmatrix}
\sqrt{\alpha_{x}} & \sqrt{1-\alpha_{x}} \\
\sqrt{1-\alpha_{x}} & \sqrt{\alpha_{x}} \\
\end{bmatrix}
\end{equation}
where $\mathrm{IL}_\mathrm{lin}$ accounts for the insertion losses and $\alpha_{x} = \frac{\mathrm{CT}}{1+\mathrm{CT}}$ represents the crosstalk, where $\mathrm{CT}$ is the ratio of the power in the undesired output port to the power in the "correct" output port (along the same propagation direction of the input).

\subsection{Comparison of interferometer architectures}

\begin{figure}[htbp]
    \centering
    \includegraphics[width=\linewidth]{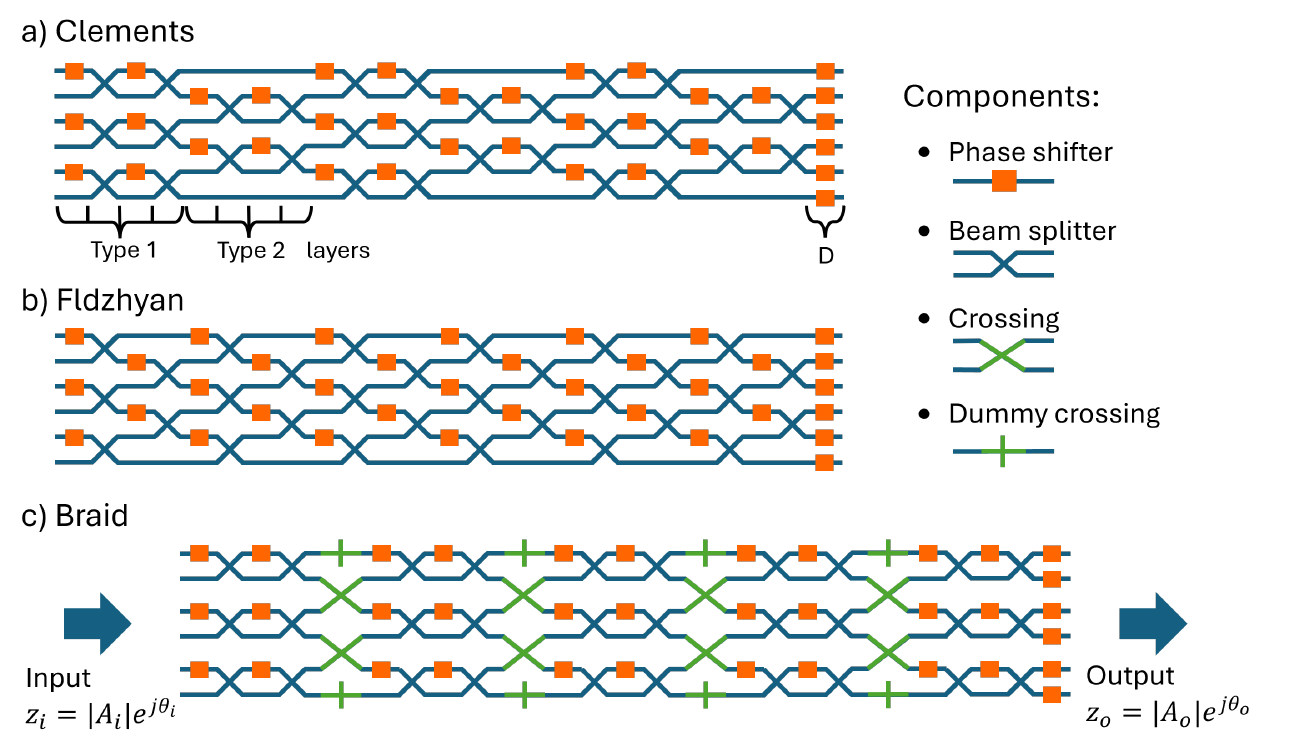}
    \caption{Schematics of the three interferometer architectures under analysis for a 6x6 matrix. (a) Clements architecture \cite{clements_optimal_2016}, with highlights of Type 1, Type 2, and $D$ layers. (b) Fldzhyan architecture \cite{fldzhyan_optimal_2020}, offering improved fabrication tolerance. (c) Our proposed braid architecture.}
    \label{fig:architectures}
\end{figure}

Figure \ref{fig:architectures} shows the three different architectures that we will compare in this paper: a) the commonly used rectangular MZI mesh commonly referred to as the Clements design \cite{clements_optimal_2016}; b) a more general mesh design of multiport interferometer as reported by Fldzhyan \cite{fldzhyan_optimal_2020} and c) our proposed braid mesh architecture. Several alternative architectures have been proposed in the literature for MVM purposes \cite{Mojaver2023_OE, Hamerly2024_X, Feng2022_ACSPhotonics, giamougiannis_coherent_2023}. However, as this work focuses on presenting the braid architecture, we have chosen to compare it with the two architectures that most closely match its performance characteristics.

We will limit ourselves to configurations with an even matrix dimension. In this case, each layer connects all input lines to a beam splitter, resulting in higher symmetry and one fewer layer compared to the Clements and Fldzhyan architectures. However, for odd matrix dimensions, one line remains unconnected, which brings the total number of layers in line with the Clements and Fldzhyan designs. As a result, despite the structural similarities, the additional crossing layers lead to worse performance compared to these architectures.

We also distinguish Type 1 layers (where only the bottom connection of each layer is a simple waveguide) and Type 2 layers (where both top and bottom connections are simple waveguides).

The braid architecture concept involves reorganizing the photonic components into a different mesh. To achieve this, crossing components are used as braids to connect each input signal to each output. This architecture enhances performance through improved symmetry, particularly in cases where the components are non-ideal. To better understand this, it is useful to analyze the different paths, such as the possible connections between any input and output.

The Clements architecture (Figure \ref{fig:architectures}.a) has certain paths that pass through fewer beam splitters, e.g., those at the very top, because of the presence of Type 2 layers. The same is true for the Fldzhyan architecture, which uses the same components as the Clements architecture but arranges them differently. In contrast, in the braid architecture, all paths pass through the same number of passive components since it only contains Type 1 layers. This results in a more balanced power flow. We added dummy crossings on both sides to maintain this balance in the crossing layer. Given the very low losses of the crossings, these dummy components are not essential and can be omitted.

We can observe that none of the three architectures maintain perfect path symmetry when considering the distribution of phase shifters per path due to the presence of the lower connection without any phase shifters. However, the braid architecture still appears to have a more uniform number of phase shifters across the paths than the other two.

A recent study by Bell et al. \cite{bell_further_2021} introduces a minor adjustment to the phase shifter in the Clements and Fldzhyan design, enhancing the balance of path insertion losses tied to the phase shifters and, consequently, improving performance. Although this modification was not applied to the braid architecture in our work, it could potentially contribute to improved path insertion loss balance for the braid design as well.

Moreover, we can define the depth of the architecture as the number of phase shifter layers encountered along the cascade (excluding the layer $D$). In the Clements and Fldzhyan architectures, the depth is $2n$. In contrast, the braid architecture has a path depth of $2(n-1)$.

Thus, we can conclude that the braid architecture offers greater symmetry in the number of components per path and a slightly lower depth, making it more robust against fabrication variations.

Finally, let us mention that it is possible to show that any unitary complex matrix $U$ requires $n^2$ real-valued degrees of freedom. Although all three architectures indeed have $n^2$ phase shifters, this alone does not directly guarantee that any unitary matrix can be obtained. Therefore, in the next Section \ref{sec:FidelitySimulation}, we will describe an optimization methodology to investigate how well an architecture can implement different unitary matrices.

Based on the previous component definitions, the full system matrix for the Braid architecture for an $N \times N$ matrix in Figure~\ref{fig:architectures} can be written as:

\begin{equation}
U = DC^{-1}_{N-1}\left( \prod_{k=N-1}^{1} C_k M_k P_k \right)
\label{eq:full_transfer_matri}
\end{equation}
with the $N \times N$ matrices $P_k$, $M_k$, and $C_k$ representing the phase shifter matrix (with respect to the first phase shifter present in odd channels), the MZI matrix and the crossing matrix, respectively. The matrix $D$ is a diagonal matrix representing an array of phase shifters at the end of the full mesh.
In the following, we define the various elements of the complete matrix of the system, where we dropped the subscript $k$ for ease of notation.
\\
The elements of the phase shifter matrix $P$ are given by: 
\begin{equation}
P_{i,j} = \sqrt{\mathrm{IL}_{PS,\text{lin}}}\,\left( e^{j\phi_{i,j}}\delta_{i,j,odd} + \delta_{i,j,even}\right)
\label{eq:shifter_matrix}
\end{equation}
where $\delta_{i,j,odd/even}$ is the Kronecker delta that holds for $i$ being an odd/even integer between 1 and $N$.
\\
The elements of the MZI matrix $M$ are given by:
\begin{equation}
\begin{aligned}
M_{i,j} = \mathrm{IL}_{\text{BS},\text{lin}} \sqrt{\mathrm{IL}_{\text{PS},\text{lin}}} \bigg[ \,
& \left( \left(\alpha_{\text{BS}} + \dfrac{1}{2}\right) e^{j\theta_{i,j}} - \left(\alpha_{\text{BS}} - \dfrac{1}{2}\right) \right) \delta_{i,j,\text{odd}} \\
+ \, & \left( \left(\alpha_{\text{BS}} + \dfrac{1}{2}\right) - \left(\alpha_{\text{BS}} - \dfrac{1}{2}\right) e^{j\theta_{i,j}} \right) \delta_{i,j,\text{even}} \\
+ \, & \sqrt{ \left( \alpha_{\text{BS}} + \dfrac{1}{2} \right) \left( \alpha_{\text{BS}} - \dfrac{1}{2} \right) } \left( e^{j\theta_{i,j}} + 1 \right) \delta_{i,j\pm1} \, \bigg]
\end{aligned}
\end{equation}
\\
The elements of the crossing matrix $C$ are given by:
\begin{equation}
\begin{aligned}
C_{i,j \,|\, i \neq 1,N} & = \sqrt{\mathrm{IL}_{C,\text{lin}}} \left( \sqrt{\alpha_{C}}\,\ \delta_{i,j}
+ \sqrt{1 - \alpha_{C}}\,\left(\delta_{i,i+1, even}
+ \delta_{i,i-1, odd}\right) \right) \\
C_{i,j \,|\, i = 1,N} & = \sqrt{1 - \alpha_{C}}\,\ \delta_{i,j}
\end{aligned}
\end{equation}
\\
The elements of the pure phase shifter matrix $D$ are given by:
\begin{equation}
D_{i,j} = e^{j\eta_{i,j}}\delta_{i,j}
\end{equation}


\section{Fidelity optimization methodology}
\label{sec:FidelitySimulation}

The main objective of our simulations is to determine the correct values of the phase shifts in a photonic architecture to approximate an arbitrary unitary matrix $U_0$. This is done using an ad-hoc PyTorch \cite{paszke2019pytorch} model for the optical circuit, which constructs its transfer matrix $U$. Its phase shifts are then optimized through backpropagation to approximate $U_0$. We hypothesize that if our new architecture is intrinsically capable of replicating any unitary matrix, such an optimization algorithm would consistently be able to tune the phase shifters to match the target matrix $U_0$, aside from minor numerical errors. Although this numerical approach cannot prove mathematically that the braid architecture can implement any unitary matrix, it can still be a practical benchmark. We will now delve deeper into the different steps of the simulations.

We employ Monte Carlo sampling to randomly sample the $U(n)$ space. This approach allows us to control the number of selected matrices while providing a more representative and diverse sampling than other methods, such as quadratic grid sampling. Our study considered $1000$ random unitary matrices, as this number is a good trade-off between computational feasibility and obtaining a statistically significant dataset. To ensure that the matrices are uniformly distributed over the unitary group $U(n)$, we employ the Haar-random unitary complex matrix generator \cite{lundberg_haar_2004}.

As shown in Equation \ref{eq:unitaryMatrixDecomposition}, the circuit transfer matrix $U$ is obtained by multiplying the matrices of all the layers of the photonic architecture, where the only way to modify $U$ is by tuning the phase shifters. Equation \ref{eq:unitaryMatrixDecomposition} represents a function that establishes a connection between two spaces: from the space of phases in the PhSs denoted by $\mathbb{R}^{n \times n}$,  to the space of unitary matrices $U(n)$ of dimension $n \times n$. The analytical inverse of this relationship is not straightforward to derive. While Clements \emph{et al.} have proposed an analytical method for their specific architecture \cite{clements_optimal_2016}, no such method has been reported for the Fldzhyan or braid architectures. While state-of-art research has explored tuning algorithms to compensate for non-ideal conditions in Clements interferometers \cite{bandyopadhyay_hardware_2021, hamerly_accurate_2022}, existing methods primarily address imbalances in the beam splitters. However, numerical optimization methods may not be optimal for Clements interferometers. We hypothesize that by leveraging backpropagation, with a sufficient number of epochs to ensure convergence, along with additional repetitions and a final set of randomly sampled points in the unitary space, performance could closely approximate the results achieved by tuning algorithmic methods. Moreover, using this numerical optimization approach across all three architectures allows for consistent and fair performance comparisons. 

Additionally, Hamerly et al. \cite{hamerly_asymptotically_2022} demonstrate that minor modifications to the Mach-Zehnder modulator’s architecture can significantly enhance imbalance tolerance. Applying similar strategies to braid architectures could yield further improvements, which we suggest as a direction for future research.

To evaluate how similar the architecture matrix $U$ is to the target matrix $U_0$, we could consider the distance in the complex matrix space $\mathbb{C}^{n \times n}$ by using the Frobenius norm: $d_F(U, U_0) = \| U - U_0 \|_F = \sqrt{\sum_{i=1}^{n} \sum_{j=1}^{n} |u_{ij} - u_{0 \: ij}|^2}$. This would not be the best approach for our purposes, as any discrepancy between $U$ and $U_0$ that involves only a scaling factor can easily be compensated. However, the Frobenius norm severely penalizes this scaling factor, making it necessary to find another metric that is insensitive to scaling. For this, we can use the so-called fidelity, which is related to the square to the cosine similarity $S$ between two vectors:

\begin{equation}
S = \frac{x \cdot y}{\| x \|_F  \cdot \| y \|_F }
\end{equation}

By flattening the matrices in vectors and using the trace operator $\mathrm{tr}$  as a notational shortcut, the fidelity $F=S^2$ can be expressed by the following equation \cite{clements_optimal_2016}:

\begin{equation}
F = \frac{\mathrm{tr}(U^\dagger U_0)^2}{N \mathrm{tr}(U^\dagger U)}
\end{equation}

Here, we used the fact that the Frobenius norm of $U_0$ is equal to $N$, because of its unitary nature.

An effective way to make the architecture matrix $U$ converge toward the target matrix $U_0$ is by tuning the phase shifters using the backpropagation method. This method relies heavily on calculating the gradient across all photonic blocks. In actual devices, fabrication tolerances introduce uncertainties that prevent precise determination of each photonic block's equations, necessitating alternative methods to find the global minimum for fabricated chips. However, for the scope of this paper, which focuses on comparing architecture performance, we will assume perfect knowledge of the photonic chip and use simulations through backpropagation.

To turn the problem into a minimization problem, we use $1-F$ as a cost function. To compare the different architectures, we generated 1000 different random target unitary matrices $U_0$ and trained the architecture 5 times, each time starting from different randomly initialized PhS values. This approach increases the probability that at least one attempt will reach the global minimum. Among the 5 repeated trained matrices, we select the result with the highest fidelity to plot. We use the Adam optimizer \cite{kinga2015method} with a learning rate of 0.001. The optimization runs for 22,000 epochs when the matrix dimension is 8. We increase the number of epochs by 1,000 every time the matrix dimension increases by 1, ensuring sufficient convergence to a minimum.


\section{Analyzing the impact of non-idealities on the fidelity}
\label{sec:fidelity_analisys}

In this section, we will analyze and compare the three photonic integrated interferometer architectures in terms of fidelity. The primary goal is to assess how accurately each architecture can reproduce a random unitary matrix with a fidelity of 1 indicating a perfect reproduction of the target matrix. The analysis begins by deriving the fidelity in the case where all the components are ideal. After establishing this baseline, we systematically examine the impact of different component-level non-idealities, assessing how each one separately affects performance. However, these non-idealities typically arise from fabrication imperfections in a chip, causing multiple effects to occur simultaneously. At the end of this section, we will combine all these effects to simulate a real-world circuit by assuming realistic component imperfection values based on current PIC technology, providing a more complete picture of the overall performance.

\subsection{Fidelity with ideal components}

\begin{figure}[htbp]
    \centering
    \includegraphics[width=1.0\textwidth]{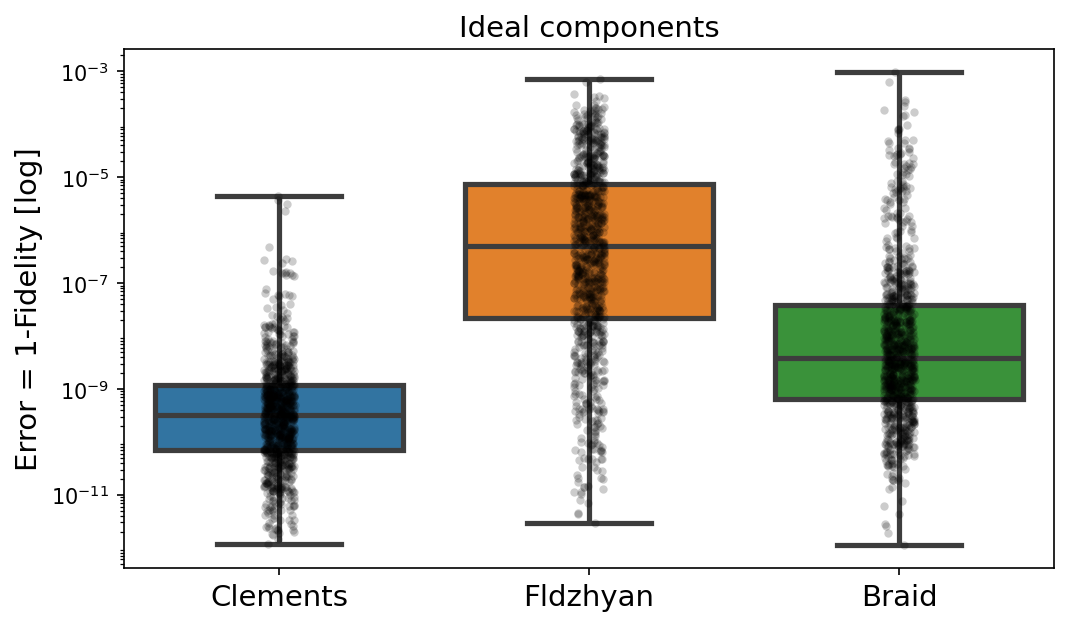}
    \caption{$\mathrm{Error} = 1-\mathrm{fidelity}$ in logarithmic scale of ideal components with 0 dB insertion loss, 0 dB imbalance, and -1000 dB crosstalk of the Clements \cite{clements_optimal_2016}, Fldzhyan \cite{fldzhyan_optimal_2020} and braid architectures. The box plot of the best out of 5 repetitions across 1000 random unitary matrices is shown. The whiskers represent the minimum and maximum values, and the points indicate the sample density. A lower value of error indicates better performance.}
    \label{fig:felityIdeal}
\end{figure}

Figure \ref{fig:felityIdeal} presents the simulation results for circuits with ideal components, with no insertion losses, imbalances, or waveguide crossing crosstalk. We display a box plot of the fidelity results for 1000 random matrices, showing the best result from 5 different random initial guesses for the phases. The results are close to perfect fidelity; therefore, for better visualization, we plot the $\mathrm{Error} = 1-\mathrm{fidelity}$ on a logarithmic scale. Additionally, the plot includes all data points for the 1000 individual matrices to give a clearer idea of the point distribution for the different architecture simulations.

Among the three architectures, Clements shows the smallest spread and the best median, indicating that it is easier to achieve convergence. Note that Clements' architecture also has a non-iterative phase decomposition algorithm that can perfectly reproduce any unitary matrix by setting the phases. However, to allow for better comparison, we opted here to also use an iterative optimization algorithm (Section \ref{sec:FidelitySimulation}). The braid architecture, despite having the same basic MZM block structure as the Clements architecture, faces slightly more challenges in convergence. We can observe that some points may get stuck in local minima or require more epochs to fully converge. Finally, the Fldzhyan architecture shows the widest point spread and worst median, with many points clustering near the maximum.

Although we have not mathematically derived an exact algorithm to represent an arbitrary unitary matrix for the braid and Fldzhyan architectures, these results demonstrate that both architectures approximate unitary matrices well, with performance close to that of the universal Clements mesh.

\subsection{Modelling non-idealities}

The decision of how to model non-idealities is challenging because, in a real chip, they are closely tied to the fabrication tolerance of the manufacturing process. Fabrication tolerances across the wafer typically show a slow gradient, for example, variation in layer thickness changes very gradually \cite{lu_performance_2017}. As a result, components that are physically close to each other are more likely to exhibit similar non-ideal characteristics \cite{dwivedi_experimental_2015}. The most accurate results should incorporate the details of these fabrication tolerances and translate them into non-ideal parameters. However, this approach is complex and requires a detailed characterization of wafer fabrication tolerances. Therefore, we will consider two extreme cases: first, all non-idealities are perfectly correlated and have the same constant value; second, all non-idealities are uncorrelated and follow a random Gaussian distribution.

The Gaussian distribution is constrained by the fact that the insertion loss must always be positive, as a negative value would imply amplification. Therefore, we use a truncated Gaussian distribution for the insertion loss, limiting the range between 0 and 2 times the average. The second truncation is necessary because an asymmetrically truncated Gaussian distribution has a mean higher than the mean of the untruncated Gaussian, and the architectures are sensitive to this shift. Since we will also sweep the standard deviation, we want to isolate its effect without simultaneously introducing a change in the expected value.

To generate the graphs in the following sections, we generated 1000 unitary matrices and repeated the simulations 5 times for each matrix, starting from different random initial phase-shifter conditions to avoid being trapped in local minima. For each simulated matrix, we calculated the corresponding fidelity values and selected the best value from the five repetitions. Using these best fidelity values, we plotted the median along with upper and lower bars representing the first and third quartiles, similar to a box plot. This approach excludes extreme values, which may result from fortunate matrix convergence or issues like optimizing around local minima. Instead, it focuses on providing an estimate that reflects most of the fidelity results.

\subsection{Impact of beam splitter non-ideality}

\begin{figure}[htbp]
    \centering
    \includegraphics[width=1.0\textwidth]{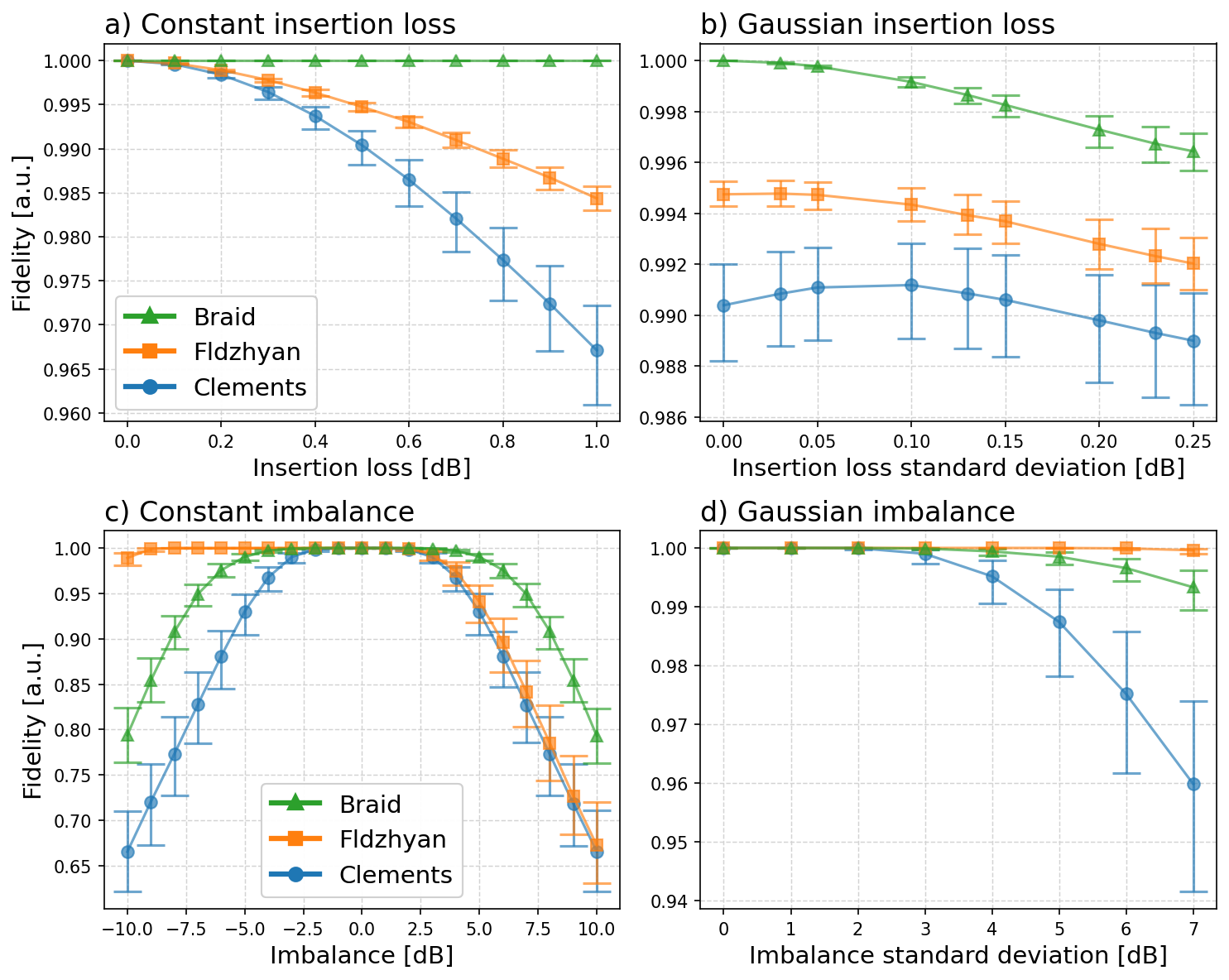}
    \caption{Fidelity results showing the median and interquartile range for different beam splitter non-idealities. These results are calculated based on the highest fidelities obtained from five initial conditions, for 1000 square matrices of size $N$=8. (a) All beam splitters have the same losses. (b) Gaussian truncated distribution of beam splitters insertion losses with an average of 0.5 dB. (c) All beam splitters have the same imbalance. (d) Gaussian distribution of imbalance. Clements and Braid have an average imbalance of 0 dB, while  Fldzhyan has average imbalance of -3.5 dB.}
    \label{fig:BS_non_ideality}
\end{figure}

Figure \ref{fig:BS_non_ideality}.a shows how fidelity is affected by sweeping the insertion loss, in the case of imperfections that are constant across the network. The braid architecture is highly robust against this effect. This is because the beam splitters are symmetric in each layer, resulting in only balanced path losses. These losses do not impact fidelity, as they only result in a scaling factor. In contrast, the Clements and Fldzhyan architectures experience unbalanced path losses due to the side paths in the Type 2 beam splitter layers (Figure\ref{fig:architectures}), which negatively affect fidelity. However, the Fldzhyan architecture appears to perform better than the Clements one. This is probably because, in the first half of the structure, the power is split evenly across all lines and recombined in the second half, therefore distributing the power more evenly and creating a better balance compared to the Clements architecture.

The resilience of the constant insertion losses in the Clements and Fldzhyan architecture can be improved by introducing a loss-matching device on the Type 2 side path to equalize path losses. This approach enhances overall performance by providing balanced loss across paths. However, adding this component will also increase insertion losses. Furthermore, since these loss-matching devices are structurally distinct from beam splitters, they may introduce additional uncertainty in actual loss values, even when designed to match specific insertion losses. A more optimal solution could involve the use of active loss-matching devices. These active components allow for tunable losses, offering greater precision in balancing the circuit’s loss profile. However, active devices often have a larger footprint, which could limit their practicality in highly integrated photonic circuits.

Figure \ref{fig:BS_non_ideality}.b shows the case where the insertion loss follows a truncated Gaussian distribution with an average of 0.5 dB, with a sweep of the standard deviation $\sigma$. The value on the left at $\sigma=0$ corresponds to a constant distribution, as in Figure \ref{fig:BS_non_ideality}.a. As we increase the standard deviation, unbalanced path losses are introduced within the architecture, leading to a drop in fidelity. For the Clements architecture, and to a lesser extent for the Fldzhyan architecture, a slight increase in the standard deviation results in a small improvement in fidelity. This could be due to a compensating effect between the path losses at the edges and the center of the architecture, which slightly benefits performance. However, this effect is not sufficient to counterbalance unbalanced path losses, which the braid architecture is not sensitive to.

In Figure \ref{fig:BS_non_ideality}.c, we vary the beam splitter imbalance, when all components have the same imbalance. In this case, more power is directed to the upper ($IMB>0$) or lower ($IMB<0$) part of the circuit, reducing the extinction ratio of the outputs and causing a drop in fidelity. As expected, the Clements and braid architectures are symmetric concerning the 0 dB imbalance (i.e., a 50/50 power split). The braid architecture performs better due to the more symmetrical positioning of the beam splitters and having fewer layers to them. On the other hand, the Fldzhyan architecture, as previously noted in \cite{fldzhyan_optimal_2020}, has an asymmetrical shape. We can derive a range of imbalances where the fidelity remains higher than 0.99: for the Clements architecture, this occurs up to 6 dB, for the braid up to 10 dB, and for Fldzhyan up to 13 dB.

In Figure \ref{fig:BS_non_ideality}.d, we sweep the standard deviation of the Gaussian distribution for the imbalance, with the average centered at the point that provides the best performance according to the constant imbalance analysis, i.e. 0 dB for the Clements and braid architectures, and -3.5 dB for the Fldzhyan architecture. If the Gaussian distribution falls entirely within that band, it is reasonable to expect the circuit to perform well. Since the Fldzhyan architecture has the widest optimal band, it is less sensitive to Gaussian-distributed imbalances. However, for current PIC technology, the standard deviation of beam splitter imbalances is typically less than 1 dB. As beam splitters are further miniaturized, fabrication tolerances play a more significant role, potentially leading to greater imbalances in splitting ratios.

Although the Fldzhyan architecture exhibits improved performance for negative imbalances, where most of the power is directed to the lower part of the circuit, this requires the design of asymmetrical beam splitters. However, designing such asymmetrical beam splitters presents challenges: it requires precise fine-tuning of the geometry, which often results in higher insertion losses compared to 50/50 beam splitters \cite{zanzi_compact_2016}, leading to a drop in fidelity.

\subsection{Impact of crossing non-ideality}

\begin{figure}[htbp]
    \centering
    \includegraphics[width=1.0\textwidth]{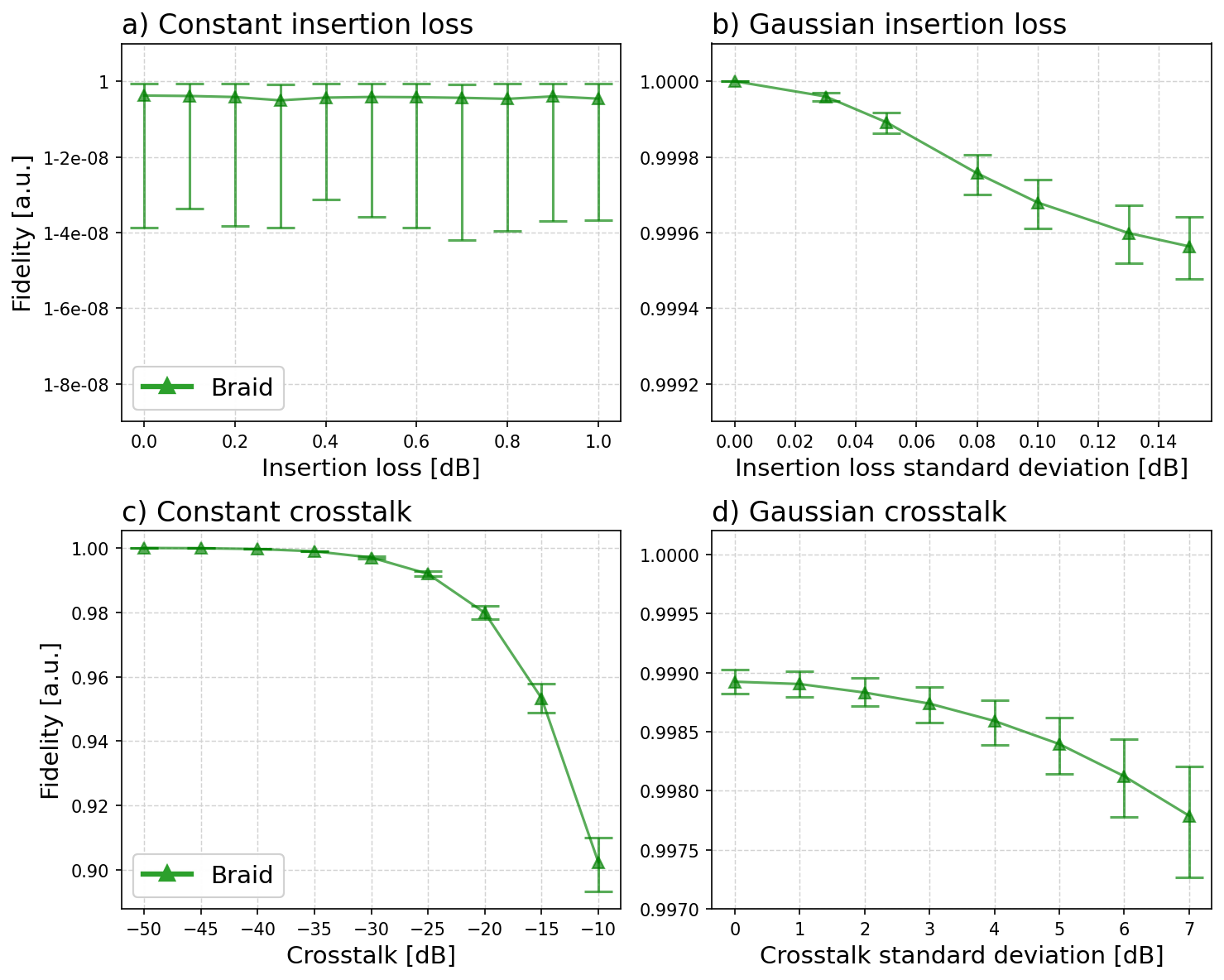}
    \caption{Fidelity braid results showing the median and interquartile range for different crossing non-idealities. These results are calculated based on the highest fidelities obtained from five initial conditions, for 1000 square matrices of size $N$=8. (a) All crossings have the same losses. (b) Gaussian truncated distribution of crossing insertion losses with an average of 0.2 dB. (c) All crossings have the same crosstalk. (d) Gaussian distribution of crosstalk with an average of -35 dB}
    \label{fig:Crossing_non_ideality}
\end{figure}

Among the architectures analyzed, the braid architecture is the only one that includes waveguide crossings. Many other types of architecture also require waveguide crossing components to function \cite{Feng2022_ACSPhotonics, giamougiannis_coherent_2023}. These components are typically considered ideal due to their remarkably good performance \cite{johnson_low-loss_2020}. However, as we will show, considering imperfections in this class of components is important, as they can still have an impact on the overall fidelity of the architecture.

Figure \ref{fig:Crossing_non_ideality}.a shows the effect of constant waveguide crossing insertion loss in the braid architecture. Since the crossings are uniformly positioned in each layer, they introduce only balanced path losses that do not degrade the fidelity. In fact, we specifically added dummy crossings on the side of our braid mesh design to maintain equal loss across all paths. This solution is straightforward, as it reuses the same crossing components, which exhibit consistent performance with the other crossings. If these dummy crossings were removed, we would observe a slight drop in performance.

In contrast, Figure \ref{fig:Crossing_non_ideality}.c shows that when the loss of insertion of a crossing has a Gaussian distribution, the varying unbalanced path losses cause a decrease in fidelity. We selected a relatively large average insertion loss of 0.2 dB, allowing us to increase the standard deviation and observe a visible impact on fidelity. However, in reality, the losses for crossings are much lower. The slope of the fidelity decline appears to converge toward a constant value. This effect arises from the truncation of the Gaussian distribution. When the standard deviation becomes too large, the truncated distribution resembles a uniform distribution.

Another important effect to consider is the waveguide crossing crosstalk. This compromises the architecture's structure, as it effectively introduces new signal paths, reducing the power in the intended paths and thereby causing unbalanced path losses. In Figure \ref{fig:Crossing_non_ideality}.b, we can see that increasing the cross-talk significantly decreases the performance. However, most crossings typically exhibit cross-talk levels lower than -35 dB \cite{johnson_low-loss_2020}, where the architecture has very good performance.

When we apply this average value to the Gaussian cross-talk distribution, we observe in Figure \ref{fig:Crossing_non_ideality}.d that the initial part of the graph shows a slow decline, indicating that the architecture is not highly sensitive to the crosstalk values following the Gaussian distribution.

As we saw from this analysis, the braid architecture is more sensitive to waveguide crossing cross-talk than to insertion loss. Therefore, it might be worth exploring a design trade-off to use crossings with higher insertion loss but lower crosstalk to improve performance.

\subsection{Impact of phase shifter non-ideality}

\begin{figure}[htbp]
    \centering
    \includegraphics[width=1.0\textwidth]{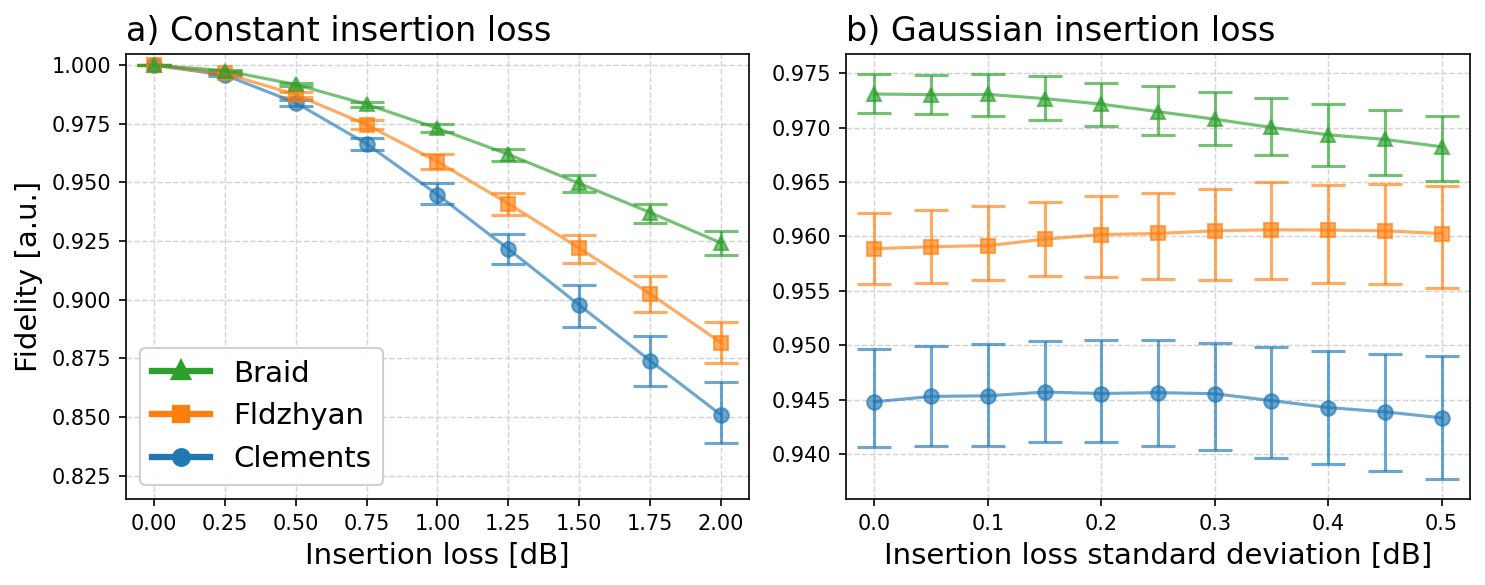}
    \caption{Fidelity results for different phase shifter non-idealities. These results are calculated based on the highest fidelities obtained from five initial conditions, for 1000 square matrices of size $N$=8. (a) All phase shifters have the same losses. (b) Gaussian truncated distribution of phase shifter insertion losses with an average of 1 dB.}
    \label{fig:PhS_non_ideality}
\end{figure}

Many emerging phase-changing technologies, such as electro-optical phase shifters or PCMs, still face the challenge of high insertion loss \cite{witzens_high-speed_2018, shafiee_compact_2023}. Therefore, it is crucial to develop architectures that are robust against these losses.

In Figure \ref{fig:PhS_non_ideality}.a, we can observe that all three architectures are significantly affected by the constant insertion loss of the phase shifters. This is because the phase shifters are positioned in each layer, and their alternating placement creates unbalanced path losses. The Fldzhyan architecture performs slightly better than the Clements architecture because of its more uniform power distribution within the circuit. However, the braid architecture demonstrates the best performance. This can be attributed to two reasons. First, it has one layer less compared to the other two architectures. Second, the placement of its phase shifters is more uniform, as it lacks the Type 2 layers where an extra connection at the bottom of the architecture is left without a phase shifter, therefore creating unbalanced paths.

In Figure \ref{fig:PhS_non_ideality}.b, we also attempt to visualize the effect of a Gaussian distribution with an average insertion loss of 1 dB. The dominance of the constant effect from the average insertion loss becomes apparent as the architecture performs relatively consistently.

\subsection{Combined analysis of non-idealities}

\begin{figure}[htbp]
    \centering
    \includegraphics[width=1.0\textwidth]{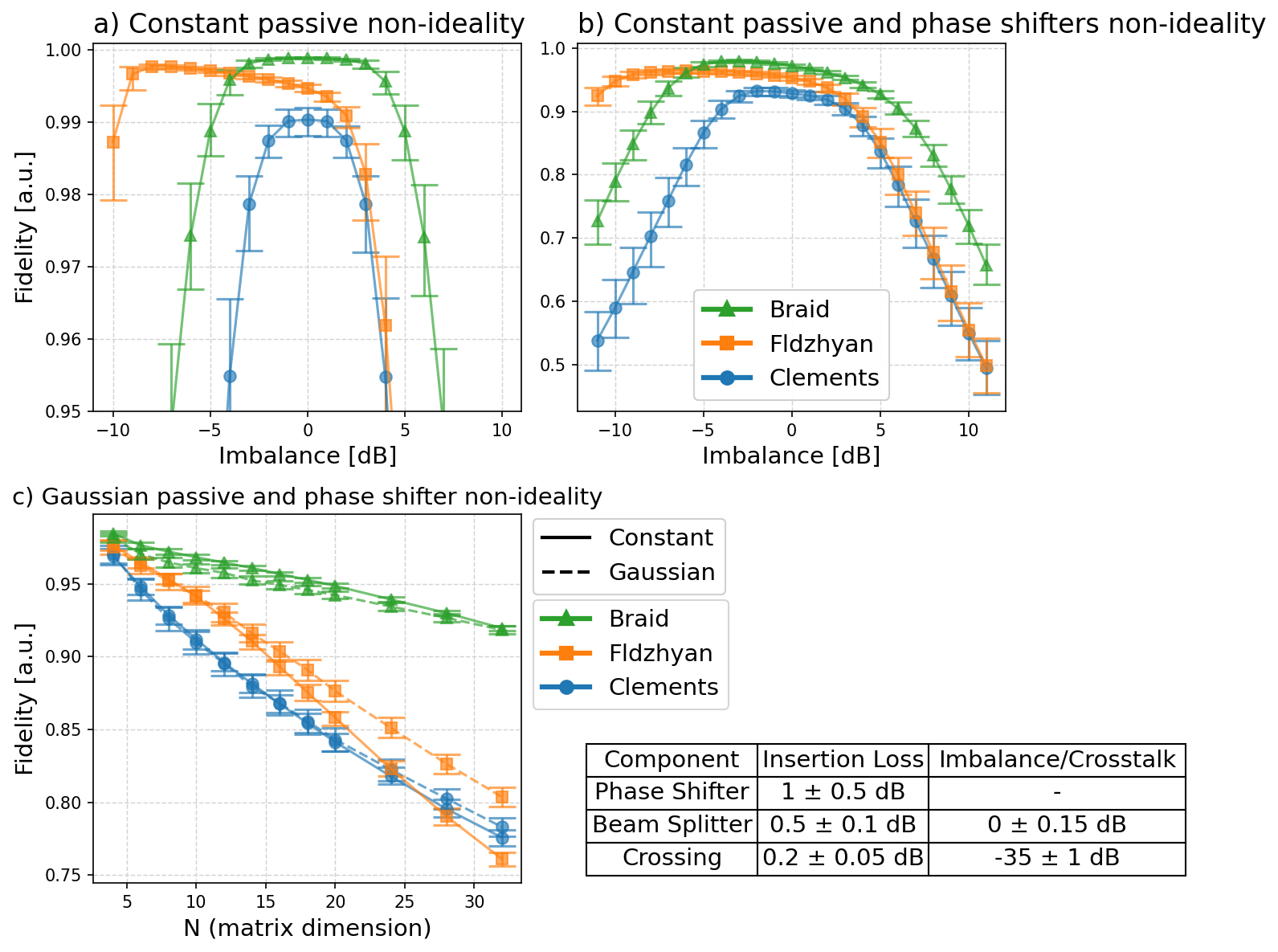}
    \caption{Fidelity results of all the combined non-ideality effects. These results are calculated based on the highest fidelities obtained from five initial conditions, for 1000 matrices (a) Constant effects of the passive devices for square matrices of size 8. The values are the averages reported in the table. (b) Constant effects of both passive and phase-shifter devices of size 8. The values are the averages reported in the table. (d) Sweep of the matrix size $N$ with constant and Gaussian distributions. The values are those reported in the table.}
    \label{fig:combined_non_ideality}
\end{figure}

For the final fidelity simulation, we combine the effects of all the imperfections.

In Figure \ref{fig:combined_non_ideality}.a, only non-ideal constant effects are considered. We sweep the phase shifter insertion loss and keep the other parameters at the average values listed in the inset. These values are inspired by realistic components, albeit with a slight overestimation of the non-idealities. The primary sources of fidelity degradation for the Clements and Fldzhyan architectures are the insertion losses of the beam splitters. In contrast, for the braid architecture, the main degrading effect arises from the crosstalk between the crossing components. However, it still outperforms the other architectures for modest values of the phase shifter insertion loss.

In Figure \ref{fig:combined_non_ideality}.b, we sweep the beam splitter imbalance and keep the other parameters at their default values. All architectures demonstrate improved performance at a negative imbalance. Therefore, directing a bit more power to the lower output of the beam splitters seems to better balance the path losses, resulting in a slight performance improvement. The superposition of all these non-ideal effects is difficult to predict, which is why simulating their combined impact is crucial.

Finally, in Figure \ref{fig:combined_non_ideality}.c, we sweep the matrix dimension, both for the case of constant imperfections and for Gaussian imperfections. Increasing the size of the interferometers results in more layers being placed in the cascade, which adds additional non-ideal effects. This figure shows that, unlike the other architectures, the braid architecture performs significantly better, especially when imperfections are constant. More importantly, because of its symmetry, it also shows superior performance as the interferometer size increases.


\section{Footprint}
\label{footprint}

A coarse method to estimate the footprint involves counting the components and multiplying that number by the footprint of each component. Table \ref{tab:numComponents} outlines the component count for each architecture. 

\begin{table}[h]
\centering
\begin{tabular}{|c|c|c|c|}  
\hline
\textbf{Architecture} & \textbf{\# PhS} & \textbf{\# BS} & \textbf{\# Crossing} \\  
\hline
Clements    & $n (n-1)$ & $n (n-1)$ & 0 \\
Fldzhyan    & $n (n-1)$ & $n (n-1)$ & 0 \\
Braid       & $n (n-1)$ & $n (n-1)$ & $(n^2-4)/2$ \\
\hline
\end{tabular}
\caption{Number of components and layer path depths for the different architectures}
\label{tab:numComponents}
\end{table}


All three architectures share the same number of phase shifters and beam splitters, but only the braid architecture includes crossings. This suggests that the braid architecture could have a larger footprint. However, in practice these crossings have a very small footprint \cite{johnson_low-loss_2020}, so it is likely that the different architectures occupy approximately the same total area.

\section{Insertion loss}
\label{insertion_loss}

Each architecture can be divided into layers, as described in Section \ref{sec:theory}. Since the layers are arranged in a cascade, the total insertion loss is calculated by determining the loss for each layer and multiplying the results. For simplicity, we assume that power is evenly distributed across all connections within a layer. Based on this assumption, we can calculate the average insertion loss for each layer.

The Clements and Fldzhyan architectures share the same types of layer, the only difference being their positioning. Therefore, it is possible to approximate the total insertion loss for these two architectures as follows:

\begin{equation}
\mathrm{IL} = \left(\frac{1+\mathrm{IL}_\mathrm{PhS}}{2} \right)^n  \left( \frac{\left(\frac{n}{2}+1\right)+\left(\frac{n}{2}-1\right) \mathrm{IL}_\mathrm{PhS}}{n} \right)^n  \mathrm{IL}_\mathrm{PhS}  \mathrm{IL}_\mathrm{BS}^n  \left( \frac{2+(n-2) \mathrm{IL}_\mathrm{BS}}{n} \right)^n
\label{eq:power1}
\end{equation}

The first two factors account for the Type 1 and 2 layers in the network, while an additional $\mathrm{IL}_\mathrm{PhS}$ is included specifically for the output layer. The second part similarly reflects the even and odd layers of beam splitters.

For the braid architecture, we can derive the following equation:

\begin{equation}
\mathrm{IL} = \left(\frac{1+\mathrm{IL}_\mathrm{PhSlin}}{2}  \right)^{2(n-1)} \mathrm{IL}_\mathrm{PhS} \: \mathrm{IL}_\mathrm{BS}^{2(n-1)}  \mathrm{IL}_\mathrm{crossing}^{(n-2)}
\label{eq:power2}
\end{equation}

The final factor accounts for the crossing layer. We intentionally exclude losses from cross-talk at dummy crossings, as these are considered negligible.

In a real-world scenario, a constant power distribution across paths in a layer is not entirely realistic, as the power varies depending on the input signal and matrix configuration. However, to assess how well the assumption of constant power holds, we simulate power consumption numerically to quantify the impact of power dispersion across different configurations. We simulated 5000 circuits, representing 1000 different target matrices, each with five repetitions. For each simulated matrix, we then multiplied it by 1000 different input vectors. The insertion losses were calculated by taking the ratio of the total output power of the architecture to the total input power. The simulation matrices were derived using the same parameters as in previous simulations, assuming constant fabrication insertion loss.

\begin{figure}[htbp]
    \centering
    \includegraphics[width=1.0\textwidth]{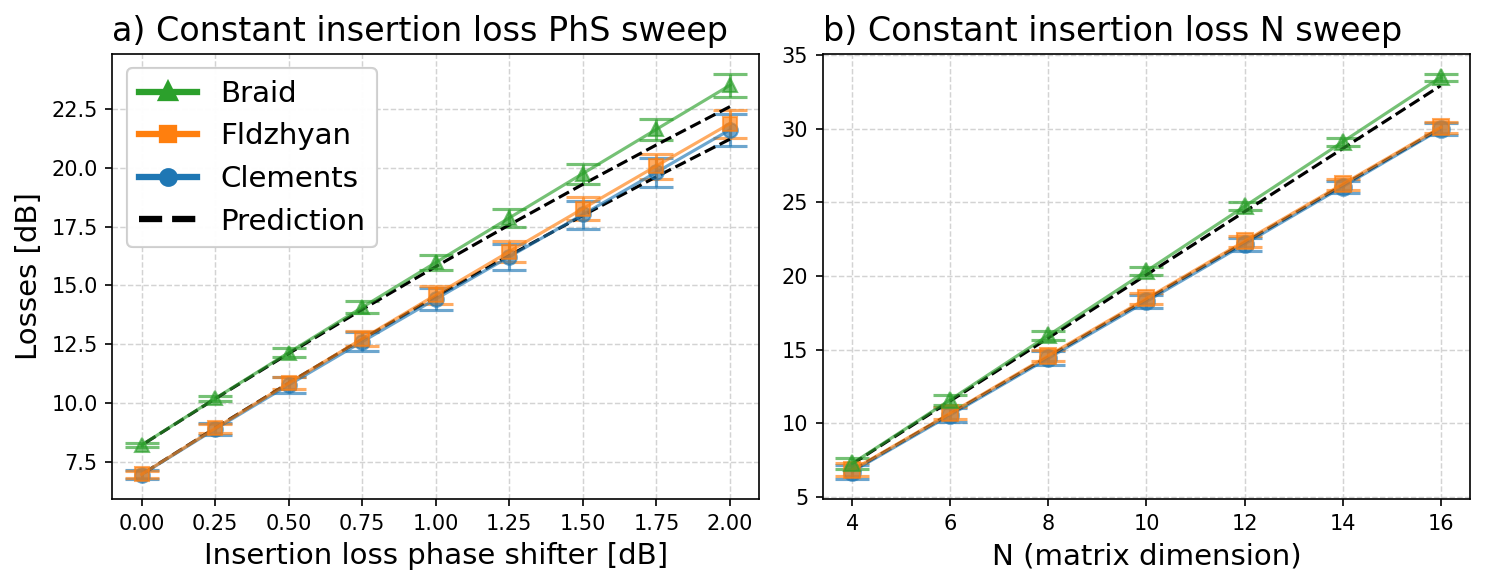}
    \caption{Simulated architecture insertion losses alongside theoretical predictions from previous equations. (a) The sweep of the constant phase shifter insertion losses, with beam splitter insertion losses of 0.5 dB, no imbalance (0 dB), crossing losses of 0.2 dB, crosstalk of -35 dB and square matrices of size 8. (b) The sweep of matrix dimension $N$, using the same device non-idealities as in (a), with constant phase shifter insertion losses of 1 dB.}
    \label{fig:powerLoss}
\end{figure}

Figure \ref{fig:powerLoss} presents the results of the numerical power simulations, using the same median and first and third quartiles as in Section \ref{sec:fidelity_analisys}.

The black dotted line represents the expected values derived from the previous approximate Equations \ref{eq:power1} and \ref{eq:power2}. As anticipated, increasing the phase shifter insertion loss or the matrix size results in higher losses within the circuit. Although we know that power is not evenly distributed across each layer, the analytical equations still provide a good approximation for most input/matrix configurations.

As expected, the braid architecture exhibits slightly higher losses compared to the other two architectures due to the additional insertion losses introduced by the crossing components. However, with advances in crossing technology \cite{johnson_low-loss_2020}, the insertion losses of these components have been reduced, narrowing the power difference between the braid architecture and the others.


\section{Future work}

In this section, we outline potential future work and additional analyses that could enhance the comparison of different architectures. It is important to note that this paper serves as an introduction to the braid architecture. Our goal was to avoid overburdening the paper while focusing on the most critical aspects. We prioritized the analysis of the most significant non-idealities and made approximations for others that we believe have a lesser impact on overall fidelity performance. Consequently, some comparisons with alternative architectures, inclusion of additional non-idealities, and evaluations of different decompositions in relation to other algorithms for matrix approximation were omitted in this initial study.

Another approach to reduce sensitivity to non-idealities involves slight modifications to the architecture. One of the simplest methods is described in the Bell paper \cite{bell_further_2021}, which introduces a minor adjustment to the phase shifters in the Clements setup. This modification effectively nullifies the fidelity sensitivity to phase shifter insertion loss. While this technique could, in principle, be applied to the braid architecture as well, we did not explore it in this study.
Other techniques, such as the use of 3-splitter MZI, offer additional benefits, as discussed in \cite{hamerly_asymptotically_2022}. While such modifications could potentially be implemented in both the Clements and braid architectures, applying them to the Fldzhyan design would be more challenging.
While these modifications could potentially be applied to the braid architecture, we opted to exclude them to ensure a more consistent comparison between the basic forms of different architectural structures.

In addition, several other architecture designs have been proposed over time, such as the product vector matrix architecture \cite{giamougiannis_coherent_2023} and the n-dimensional beam splitters as reported in \cite{tanomura_robust_2020}. Also, diamond-based architectures, such as the diamond mesh, a phase-error- and loss-tolerant field-programmable MZI-based optical processor for optical neural networks \cite{shokraneh_diamond_2020}, have also been proposed. These various architecture types could also be compared in future work.

In our analysis, we did not account for the phase differences in both the waveguides and, more importantly, in the building blocks. Specifically, the phase shifts between the outputs of the MMIs and crossings are not guaranteed to remain constant. These phase shifts could vary, thereby altering the interference between different paths, which in turn influences the overall performance of the circuit. Additionally, simulating the quantization effects of phase shifts may also be necessary. While we consider these effects to be less significant, additional simulations are still needed to provide a fuller picture.

In the ideal Clements architecture \cite{clements_optimal_2016}, we can derive the exact mathematical mapping of the selected weight matrix using a prescriptive algorithm. Even in the presence of non-idealities, \cite{hamerly_accurate_2022} demonstrates how a robust mathematical solution can be derived. It would be valuable to investigate whether, in such cases, the Clements architecture still underperforms compared to other architectures. However, since neither the Fldzhyan nor the braid architecture has a direct mathematical algorithm for reproducing the target matrix, such a comparison is not feasible. That said, the backpropagation optimization we employed should converge to the optimal solution. Further research is needed for both the Fldzhyan and braid architectures to develop a mathematical framework that links the target matrix to the phase shifter space.

\section{Conclusion}

In this paper, we proposed a novel photonic interferometer braid architecture capable of approximating any complex unitary matrix with high accuracy in terms of fidelity. Our architecture offers a substantial performance improvement in the presence of non-idealities when compared to state-of-the-art SVD-based architectures, specifically the Clements and Fldzhyan designs. Through detailed numerical simulations, we investigated the ideal case and then systematically introduced different non-ideal components, including insertion loss, beam splitter imbalances, phase shifter imperfections, and crosstalk, to assess the robustness of each architecture.

As mentioned before, a recent paper by Bell et al. \cite{bell_further_2021} presents a slight modification to the phase shifter in the Clements and Fldzhyan design, which improves the balance of path insertion losses associated with the phase shifters and therefore the fidelity performance. While we did not apply Bell’s modification to the braid architecture, in practice, it could further increase the path insertion loss balance in the braid design as well.

When introducing non-idealities, different architectures responded differently. The braid architecture exhibited strong robustness to insertion losses and beam splitter imbalances, thanks to its symmetrical design and fewer layers, which balanced the losses more evenly. However, it is more susceptible to crosstalk from crossings, though this effect is minimal for typical low-crosstalk values. The Fldzhyan architecture performed well under beam splitter imbalance conditions, but to achieve optimal fidelity performance, imbalanced beam splitters would have to be used. These beam splitters are typically more lossy because of their asymmetry, which ultimately reduces the overall fidelity performance. The Clements architecture, though optimal under ideal conditions, showed a significant drop in performance when affected by non-ideal components.

Finally, in a real-world scenario, where all non-idealities were combined, the braid architecture emerged as the most robust, particularly as the size of the interferometer increased. Its symmetrical design made it more resistant to the cumulative effects of imperfections. As interferometer dimensions grew, the braid architecture's superior fidelity performance became more apparent, outperforming the Clements and Fldzhyan architectures, especially under constant non-idealities.

In addition to fidelity, we also evaluated the footprint and insertion loss for the three architectures. Although all architectures share the same number of phase shifters and beam splitters, the braid architecture includes waveguide crossings, which slightly increase its overall footprint. In terms of insertion loss, the braid architecture exhibited marginally higher losses due to additional crossings. However, with advancements in crossing technology, these components now have a very small footprint and low insertion loss, making the differences in both footprint and power losses negligible.

As we scale to larger circuits, miniaturization inherently brings more non-idealities, such as increased insertion loss, crosstalk, and phase shifter imperfections. In this context, the braid architecture stands out as a robust interferometer design, offering the ability to maintain meaningful results despite these imperfections. By comparison, the Clements and Fldzhyan architectures, though effective under ideal conditions, struggle with robustness as circuit size and complexity increase, making them less practical for large-scale implementations. In summary, the braid architecture ensures that even with more complex and realistic designs, the circuit can still function appropriately, supporting a broader range of applications where architecture robustness is crucial.


\section*{Funding}
This work has received funding from the European Union’s Horizon
Europe research and innovation program under grant agreement
No. 101070238.

\section*{Disclosures}
The authors declare that there are no conflicts of interest related to this article.

\section*{Data availability statement}
The data that support the findings of this study are available from the corresponding author upon reasonable request.

\bibliographystyle{unsrt}  


\end{document}